\documentclass[prb,aps,twocolumn,showpacs,superscriptaddress]{revtex4}
\usepackage{amsfonts,amsmath,bm}

\usepackage{graphicx}

\newcommand{\kp}{|+\rangle}
\newcommand{\km}{|-\rangle}
\newcommand{\bp}{\langle+|}
\newcommand{\bmm}{\langle-|}
\newcommand{\ktp}{|T+\rangle}
\newcommand{\ktm}{|T-\rangle}
\newcommand{\kup}{|\!\uparrow\rangle}
\newcommand{\kdn}{|\!\downarrow\rangle}
\newcommand{\ktup}{|T\!\uparrow\rangle}
\newcommand{\ktdn}{|T\!\downarrow\rangle}
\newcommand{\bup}{\langle\uparrow\!|}
\newcommand{\bdn}{\langle\downarrow\!|}
\newcommand{\btup}{\langle T\!\uparrow\!|}
\newcommand{\btdn}{\langle T\!\downarrow\!|}
\newcommand{\pup}{|\!\uparrow\rangle\!\langle\uparrow\!|}
\newcommand{\pdn}{|\!\downarrow\rangle\!\langle\downarrow\!|}
\newcommand{\ptup}{|T\!\uparrow\!\rangle\!\langle T\!\uparrow\!|}
\newcommand{\ptdn}{|T\!\downarrow\!\rangle\!\langle T\!\downarrow\!|}
\newcommand{\wt}{\omega_{\mathrm{t}}}
\newcommand{\wh}{\omega_{\mathrm{h}}}
\DeclareMathOperator{\atan}{atan}
\DeclareMathOperator{\tr}{Tr}
\DeclareMathOperator{\im}{Im}
\DeclareMathOperator{\re}{Re}

\begin{document}

\title{Theory of the time-resolved Kerr rotation on trapped holes}

\author{Pawe{\l} Machnikowski}
 \email{Pawel.Machnikowski@pwr.wroc.pl} 
\affiliation{Institute of Physics, Wroc{\l}aw University of
Technology, 50-370 Wroc{\l}aw, Poland}
\author{Tilmann Kuhn}
\affiliation{Institut f{\"u}r Festk{\"o}rpertheorie, Westf{\"a}lische
Wilhelms-Universit{\"a}t M{\"u}nster, 48149 M{\"u}nster, Germany}

\begin{abstract}
We formulate a model of the time-resolved Kerr rotation experiment on
an ensemble of independent holes in a semiconductor nanostructure 
(e.g., confined in a quantum dot or trapped in a quantum well) in a
tilted magnetic field. We use a
generic Markovian description of the hole and trion dephasing and
focus on the interpretation of the time-resolved signal in terms of
the microscopic evolution of the spin polarization. We show that the
signal in an off-plane field contains components that reveal both the
spin relaxation rate and the spin coherence dephasing rate. We derive
analytical formulas for the hole spin polarization, which may be used
to extract the two relevant rates by fitting to the measurement data.
\end{abstract}


\maketitle

\section{Introduction} 

In recent years, considerable experimental progress has been achieved
in the optical control and readout of spin states of electrons and
holes in semiconductor nanostructures
\cite{shabaev03,dutt05,greilich06c,kennedy06,atature07,greilich07,press08,gerardot08}.
Extended life times of spin states in quantum dots (QDs) \cite{paillard01}
and quantum
wells (QWs) \cite{kikkawa97} seem very promising for applications in spintronics,
e.g. in the form of spin memory \cite{kikkawa97,kroutvar04}, or in
semiconductor spin-based quantum computing
\cite{loss98,hawrylak05}. Particular expectations are related to hole
spins, since the reduced hyperfine interaction in this case makes the
hole spin decoherence even slower \cite{heiss07,brunner09}.

In order to design feasible spin-based devices, one obviously has to
understand the properties of spins in confined semiconductor
systems. The most essential parameters that must be known in order to
predict the evolution of a spin in a real structure are the Land\'e
$g$-tensor, which defines the unitary evolution of a spin in a magnetic
field, and the relaxation (dephasing) times, accounting for the impact
of the environment. As the Zeeman shifts can be rather small and,
therefore, hard to resolve spectrally, many experiments rely on
time-resolved methods \cite{bar-ad91,damen91,marie99,palinginis04}. Among
those, time-resolved Faraday \cite{tribollet03,meier07,yugova07,atature07} or
Kerr \cite{malinowski00,zhukov07,syperek07,zhukov09,kugler09} rotation
measurements have proven
to be very useful. In these experiments, one studies the rotation of
the polarization plane of transmitted or reflected radiation (probe pulse) due
to spin polarization excited by a circularly polarized pump pulse. In
this way, the decay of spin polarization and the spin precession
around the quantization axis can be observed as a function of delay
time between the pulses.

The theoretical challenge in the description of confined hole spin
properties is essentially 
three-fold. First, one has to model the nanostructure in order to find
the effective $g$-factor for holes in a two- or zero-dimensional
confinement. This is usually done using $k\cdot p$ methods
\cite{marie99,zhao94a,zhao94b} and the results are in reasonable
agreement with measurements. Second, one needs to describe the
spin dephasing and relaxation processes. Phonon-mediated processes are
often invoked for QDs \cite{woods04,bulaev05a}, although other channels
are also taken into account \cite{sanjose06}. In QWs, phonons are
expected to dominate for trapped holes, while scattering due to compositional
disorder is invoked for delocalized ones \cite{ikonic01}. Another
reason for dephasing may be system inhomogeneity, in particular
$g$-factor fluctuations \cite{semenov02}. 

In the present paper, we deal with the third aspect of the problem,
namely, the microscopic origin of the measured signal and, more
importantly, the relation between 
the magnetic (spin) orientation, which is supposed to be studied, and
the optical field, which is experimentally accessible. We analyze also
in what way the optical response (in particular, Kerr
rotation) of the system depends on the parameters characterizing the
system evolution. In the case of a time-resolved experiment with
pulsed excitation, the physical interpretation of the detected optical
response becomes non-trivial and constitutes a subject of study by
itself \cite{zhang09,yugova09}. This kind of
theoretical discussion has been presented for excitons in a QW
\cite{ostreich95} and, on a phenomenological level, for an n-doped QW
system \cite{zhukov07}.  
Very recently, a complete analysis of the Kerr and Faraday response
for ensembles of n-doped QDs in an in-plane magnetic field was
presented \cite{yugova09}.
Here, we focus on trapped hole states in QWs and on holes
confined in QDs. We discuss the microscopic origin of the
time-resolved Kerr rotation (TRKR) signal in a pump-probe experiment
on a p-doped sample in tilted magnetic field. We perform a complete
analysis in the density matrix formalism and describe
hole spin dephasing on a general level, assuming only its Markovian
character. Our description is applicable to various decoherence
processes that have a well-defined Markov limit wchich is applicable
under given conditions. Examples of such processes include
phonon-assisted transitions or Coulomb scattering. We discuss how the
``longitudinal'' and ``transverse'' 
dephasing rates (defined with respect to the tilted quantization axis)
manifest themselves in the detected TRKR signal. 

As a result of our study, we point out that the optical signal
follows the spin polarization in the limit of strong dephasing
of optical coherences. 
The same holds true in the case of long-lived optical coherences if the
phase relation between the pulses can be considered random or, for
an extended system, if the experimental geometry assures that the
coherent part of the signal is emitted in a different direction.
We show that the dephasing of the hole spin
precession beats is governed by two rates involving a combination of
the two relaxation rates. In a homogeneous system, this may allow one
to extract both rates from a single experiment. In addition, we study
the effect of inhomogeneity of $g$-factors on the observed TRKR signal.

The paper is organized as follows. In Sec. \ref{sec:system}, we
describe the system and the experiment to be modelled. The next
Sec.~\ref{sec:signal} contains the microscopic derivation of the
optical TRKR signal. In Sec.~\ref{sec:decoh}, we study the spin
dynamics using a general model of Markovian decoherence, including
also inhomogeneous dephasing. The relation between the optical signal
and the spin polarization is discussed in Sec.~\ref{sec:spin-opt}. In
Sec.~\ref{sec:results}, we present the results of our simulations and
discuss the dependence of the TRKR signal on the off-plane tilt angle
of the magnetic field and on the relative strength of different
contributions to dephasing (longitudinal, transverse,
inhomogeneous). These results are discussed in the next
Sec.~\ref{sec:discussion}. The appendix contains the derivation of the
general Lindblad equation describing Markovian decoherence in the
hole-trion spin system.

\section{The system}
\label{sec:system}

We consider the optical response of a system composed of trapped
(localized) holes which may be considered independent
(non-interacting). This situation may correspond to a p-doped quantum
well in which holes are localized by some kind of
trapping potentials, e.g., interface fluctuations or nearby
defects. The theory applies also to ensembles of quantum
dots in a remotely p-doped structure. In thermal equilibrium, each
trapping center is assumed to accommodate one hole.  
The area density of such trapped holes is $\nu$. 
Only heavy hole
states are considered, since the 
heavy-light hole splitting is usually large in confined systems.
The fundamental optical transition at each trapping center consists in
an excitation of an electron-hole pair which, together with the
resident hole, forms a bound trion. 
It is assumed that the
temperature is low and the driving pulses are spectrally narrow
enough to restrict the description to the lowest hole and trion
states. 
The system is placed in a magnetic field $\bm{B}$ oriented at an angle
$\theta$ with respect to the growth axis. 
The quantum well or
layer of QDs is covered by a capping layer of thickness $D$.

A single trapped hole-trion system is described by the Hamiltonian
\begin{eqnarray}\label{ham0}
H_{0}=-\frac{1}{2}\mu_{\mathrm{B}}\bm{B}\hat{g}_{\mathrm{h}}\bm{\sigma}_{\mathrm{h}}
-\frac{1}{2}g_{\mathrm{t}}\mu_{\mathrm{B}}\bm{B}\cdot\bm{\sigma}_{\mathrm{t}},
\end{eqnarray}
where $\mu_{\mathrm{B}}$ is the Bohr magneton,
$\hat{g}_{\mathrm{h}}$ is the hole Land{\'e} tensor, 
$g_{\mathrm{t}}$ is the Land{\'e} factor of the trion
(i.e., essentially, of the electron), which we assume isotropic, and 
$\bm{\sigma}_{\mathrm{h}},\bm{\sigma}_{\mathrm{t}}$ are the vectors of
Pauli matrices corresponding to the hole and trion spin, respectively
(the hole is treated as a pseudo-spin-1/2 system). 
Here and in the following, we describe the system in a reference
frame rotating with the zero-field hole-trion transition frequency $\omega$.

The spin states of each hole or trion can be described in terms
of the ``spin up'' and ``spin-down'' states, that is, the basis states
with definite projections on the growth axis 
(normal to the QW or to the plane of QDs), $\kup,\kdn$ (for a hole)
and  $\ktup,\ktdn$ (for a trion).
For the
magnetically isotropic trion, we define the two Zeeman eigenstates
\begin{eqnarray*}
\ktp & = & \cos\frac{\theta}{2}\ktup+\sin\frac{\theta}{2}\ktdn,\\
\ktm & = & -\sin\frac{\theta}{2}\ktup+\cos\frac{\theta}{2}\ktdn.
\end{eqnarray*}
In the case of the hole, the quantization axis does not necessarily
coincide with the field orientation. 
The hole spin eigenstates can be written as 
\begin{eqnarray*}
\kp=\cos\frac{\phi}{2}\kup+\sin\frac{\phi}{2}\kdn,\\
\km=-\sin\frac{\phi}{2}\kup+\cos\frac{\phi}{2}\kdn,
\end{eqnarray*}
where $\phi$ is a certain angle depending on the structure of the
hole Land\'e tensor. In our simulations, the latter will be
assumed isotropic in the structure plane, so that 
\begin{equation}\label{q-axis}
\phi=\atan\left[ \frac{g_{\mathrm{h}\bot}}{g_{\mathrm{h}||}}\tan\theta \right],
\end{equation}
where $g_{\mathrm{h}\bot}$ and $g_{\mathrm{h}||}$ are the in-plane and
axial components of $\hat{g}$. The Zeeman energy splitting for the
hole is then
$\hbar\omega_{\mathrm{h}}
=\sqrt{g_{\bot}^{2}\sin^{2}\theta+g_{||}^{2}\cos^{2}\theta}\mu_{\mathrm{B}}B$,
Note, however, that the
actual structure of the Land\'e tensor enters the theory only via the
the angle $\phi$ and the Zeeman energy $\hbar\omega_{\mathrm{h}}$
which can easily be found also for structures with a more complicated
form of the Land\'e tensor \cite{kusraev99}.

A laser tuned to the trion line excites the system by two pulses (pump-probe
configuration). The pulses propagate nearly perpendicular to the
structure plane (a small deviation of the probe beam from the
perpendicular axis is needed to 
separate the contributions to the third order response, as discussed
in Sec.~\ref{sec:spin-opt}).
The first pulse arrives at $t=0$ and is circularly
polarized  ($\sigma_{+}$). The second, linearly ($X$) polarized one
arrives at $t=\tau$. The amplitudes of the electric field in the two
pulses (outside the semiconductor) are $E_{i}=|E_{i}|e^{-i\psi_{i}}$,
$i=1,2$. The electric 
field couples to the interband transitions via a dipole moment
$d$. The pulse shape is described by a function $\eta(s)$, which
is of the order of unity. The pulse length will be denoted by
$\tau_{\mathrm{p}}$. The reflection amplitude at the 
semiconductor-vacuum interface is $r=(1-n)/(1+n)$, where $n$ is the
refractive index of the capping layer.
The relevant
Hamiltonian in the rotating wave approximation is then
\begin{eqnarray}\label{ham-las}
H_{\mathrm{las}} & = & \frac{A_{1}}{2}\eta\left(\frac{t}{\tau_{\mathrm{p}}}\right)
e^{i\psi_{1}}\kup\!\btup \nonumber\\
&&+\frac{A_{2}}{2\sqrt{2}}\eta\left(\frac{t-\tau}{\tau_{\mathrm{p}}}\right)
e^{i\psi_{2}}\left( \kup\!\btup +\kdn\!\btdn\right) \nonumber \\
&&+\mathrm{H.c.},
\end{eqnarray}
where $A_{i}=d|E_{i}|(1+r)$ are the effective amplitudes
of the two pulses inside the material. 
For the pulse shapes we will assume Gaussians, 
$\eta(s)=\exp[-(1/2)s^{2}]$.

In addition to the evolution governed by the Hamiltonian
$H=H_{0}+H_{\mathrm{las}}$, the system undergoes dissipative dynamics due to the
interaction with its environment. As long as this decoherence can be
described in the Markov limit (which is reasonable in view of the
relatively long time scales involved), its effect on the hole spin can be described by
the universal Lindblad superoperator (see Appendix
\ref{app:lindblad}). In this limit, the 
open system evolution is described by three dephasing rates:
$\kappa_{\pm}$ describe the
longitudinal decoherence, that is, spin relaxation between the Zeeman
eigenstates for a given field orientation, while $\kappa_{0}$ accounts
for additional pure dephasing processes. 
The two spin-flip rates $\kappa_{\pm}$ are related by
\begin{displaymath}
\kappa_{-}=\kappa_{+}\exp\left[ 
-\frac{\hbar\wh}{k_{\mathrm{B}}T} \right], 
\end{displaymath}
which guarantees the detailed balance condition at equilibrium. Here
$\kappa_{+}$ is the transition rate for a ``down-flip'' (from the
upper to the lower Zeeman 
state), and $\kappa_{-}$ is the rate for an ``up-flip''. Note that,
apart from this relation, in a specific model of hole-reservoir
interaction, the rates $\kappa_{\pm}$ will depend on the
Zeeman splitting and temperature via the spectral density of the
relevant reservoir [see Eq.~(\ref{kappa-R})]. However, in our general
discussion, we treat one 
of them (or their sum) as a parameter of the model.

The Lindblad dissipator describing the hole
decoherence has the form
\begin{eqnarray}\label{lind-hole}
L_{\mathrm{h}}[\rho] & = & \kappa_{+}\left[ 
\sigma_{-}^{\mathrm{(h)}}\rho\sigma_{+}^{\mathrm{(h)}}
-\frac{1}{2}\left\{
\sigma_{+}^{\mathrm{(h)}}\sigma_{-}^{\mathrm{(h)}},\rho\right\}_{+}
\right] \nonumber\\
&&+\kappa_{-}\left[ 
\sigma_{+}^{\mathrm{(h)}}\rho\sigma_{-}^{\mathrm{(h)}}
-\frac{1}{2}\left\{
\sigma_{-}^{\mathrm{(h)}}\sigma_{+}^{\mathrm{(h)}},\rho\right\}_{+}
\right] \nonumber\\
&&+\frac{1}{2}\kappa_{0}\left[ 
\sigma_{0}^{\mathrm{(h)}}\rho\sigma_{0}^{\mathrm{(h)}}
-\frac{1}{2}\left\{
\sigma_{0}^{\mathrm{(h)}}\sigma_{0}^{\mathrm{(h)}},\rho\right\}_{+}
\right],
\end{eqnarray}
where 
\begin{displaymath}
\sigma_{+}^{\mathrm{(h)}}=\left[ \sigma_{-}^{\mathrm{(h)}}
\right]^{\dag} = \kp\!\bmm,\quad 
\sigma_{0}^{\mathrm{(h)}}=\kp\!\bp-\km\!\bmm.
\end{displaymath}
An analogous dissipator describes the spin dephasing of the
trion. However, since the trion spin coherence time is much longer than
its lifetime, the decay of spin coherence will be governed by the
latter and the trion spin dephasing can be neglected.
Note that the spin dephasing
in the Markov limit is necessarily described in the eigen basis
$\kp,\km$ defined by 
the field orientation, which differs from the ``spin-up'' and
``spin-down'' basis defined by the structure symmetry and by the
optical selection rules. 

The last part of the model is the radiative decay of the trion, which
is accounted for by the Lindblad superoperator
\begin{eqnarray}\label{lind-radiat}
L_{\mathrm{rad}}[\rho] & = & \gamma_{1}\left[ 
\sigma_{-}^{(\uparrow)}\rho\sigma_{+}^{(\uparrow)}
-\frac{1}{2}\left\{
\sigma_{+}^{(\uparrow)}\sigma_{-}^{(\uparrow)},\rho\right\}_{+} 
\right.\nonumber \\
&&\left.+\sigma_{-}^{(\downarrow)}\rho\sigma_{+}^{(\downarrow)}
-\frac{1}{2}\left\{
\sigma_{+}^{(\downarrow)}\sigma_{+}^{(\downarrow)},\rho\right\}_{+}
\right] \nonumber \\
&&+\frac{1}{2}\gamma_{0}\left[ 
\sigma_{0}\rho\sigma_{0}-\frac{1}{2}\left\{
\sigma_{0}^{2},\rho\right\}_{+}
\right],
\end{eqnarray}
where $\gamma_{1}$ is the radiative decay rate, $\gamma_{0}$ is the
additional pure dephasing rate, and the transition operators are 
\begin{eqnarray*}
\sigma_{+}^{(\uparrow)}=\left[ \sigma_{-}^{(\uparrow)}
\right]^{\dag} = \kup\!\btup,\quad 
\sigma_{+}^{(\downarrow)}=\left[ \sigma_{-}^{(\downarrow)}
\right]^{\dag} = \kdn\!\btdn,\\
\sigma_{0}=\ptup+\ptdn-\pup-\pdn.
\end{eqnarray*}
Note that the distinction between the trion recombination and pure
dephasing is essential here not only because of the presence of
various pure dephasing mechanisms in real systems
\cite{krummheuer02,borri05,muljarov04,machnikowski06e} but, much more importantly,
because of the different effect these processes have on the
spin-dependent optical response: both of them contribute to the decay
of the optical polarization but pure dephasing, contrary to recombination,
does not affect the trion spin occupations. 
In an ensemble of emitters, the dephasing of trion coherences can be
in fact dominated by inhomogeneous effects (distribution of the trion
transition frequencies). This would result in a different form of the
coherence decay. However, from the point of view of the present study,
this difference is of minor importance and only the
characteristic time of the coherence decay is essential. Therefore, we
simplify the discussion by neglecting this kind of inhomogeneity and
using only the pure dephasing rate $\gamma_{0}$ to characterize the optical
dephasing.  

\section{The TRKR response}
\label{sec:signal}

In this Section, we define the measured TRKR signal and clarify its
relation to the microscopic variables (elements of the density matrix) 
defining the state of the carriers in a nanostructure at the moment
when the probe pulse arrives. We show how the phenomenology of
Kerr rotation emerges in the homodyne detection process from the
interference of the macroscopic optical field reflected
from the system surface with the radiation 
due to the interband optical polarization in the nanostructure. Finally, we
relate the latter to the spin polarization.

The experimentally measured effect is 
a rotation of the polarization 
plane of the probe beam reflected from the sample. 
The total field is projected
onto the two axes $x,y$, oriented at $45^{\circ}$ with respect to the
polarization of the probe beam. 
The rotation of the polarization axis is given by the difference
of intensity between the corresponding two components of the field
\cite{yugova09,zhukov07},
\begin{equation}
\Delta I=\frac{1}{\mu_{0}c}
\left[  \langle E_{y}^{2}(t)\rangle 
- \langle E_{x}^{2}(t)\rangle \right]
=\frac{1}{\mu_{0}c}\im\left( 
E_{+}E_{-}^{*} \right), 
\label{intensity}
\end{equation}
where  $E_{+}$ and $E_{-}$ 
denote the (complex) amplitudes of the circularly right-
and left- polarized components of the total field
and $\langle\cdot\rangle$ denotes time averaging over the period of
the electromagnetic field.

On the microscopic level, the observed reflected field is a sum of the
beam reflected at the surface of the capping layer (this process will
be treated on the usual, macroscopic level) and the field emitted by
the nanostructure. 
Thus, the two circular polarization components of the total
field incident at the detector are 
\begin{equation}
E_{\pm}=E_{\mathrm{R}\pm}+E_{\mathrm{S}\pm},
\label{total-signal}
\end{equation}
where $E_{\mathrm{R}\pm}$ is the field reflected from the
surface of the capping layer and $E_{\mathrm{S}\pm}$ is
the field emitted by the carries trapped in the nanostructures. 
For pulsed excitation, slow evolution of the field amplitudes has to
be taken into account. The field reflected at the surface simply
follows the pulse envelope and the
amplitudes of its $\sigma_{+}$ and $\sigma_{-}$ components at the
sample surface are both equal to 
\begin{equation}
E_{\mathrm{R}\pm}(t)\equiv E_{\mathrm{R}}(t)=\frac{1}{\sqrt{2}}
rE_{2}\eta\left(\frac{t-\tau}{\tau_{\mathrm{p}}}\right)
\label{reflected}.
\end{equation}  

The optical signal emitted from
the structure originates from the interband polarization. If
$\rho(t)$ denotes the density matrix representing the system state
then each trapped 
hole-trion superposition contributes a $\sigma_{+}$ component of the
dipole moment 
$\pi_{+}=d\btup\rho(t)\kup e^{-i\omega t}+\mathrm{c.c.}$
and a
$\sigma_{-}$ component 
$\pi_{-}=d\btdn\rho(t)\kdn e^{-i\omega  t}+\mathrm{c.c.}$.
This 
results in the polarization current 
\begin{displaymath}
\mathcal{J}_{+}=-i\nu\omega d\btup\rho(t)\kup
e^{-i\omega t}+\mathrm{c.c.}
\end{displaymath}
(and analogous for $\mathcal{J}_{-}$)
and the amplitudes 
of the corresponding two components of the radiation emitted from the
structure (at the sample surface) are
\begin{equation}\label{Epm}
\left(\begin{array}{c}E_{\mathrm{S}+}(t)\\ 
E_{\mathrm{S}-}(t)\end{array} \right)=
\frac{i}{2}\mu_{0}c\nu d\omega 
\left(\begin{array}{c} 
\btup\rho(t)\kup \\ \btdn\rho(t)\kdn\end{array}
\right)e^{-i\varphi},
\end{equation}
where $\varphi=2D\omega n/c$ is the phase shift (with
respect to the field $E_{\mathrm{R}}$) due to propagation through the
capping layer. 

Note that if the two components have equal phases (as is
indeed the case, see below) then the radiation emitted from the
nanostructure is, in 
general, elliptical but its polarization axis is not
rotated. Moreover, the intensity of this signal is weak. What one
really measures in the homodyne detection scheme is, however, the signal 
coherently superposed on the much stronger field reflected from the
surface of the sample. Substituting Eqs.~(\ref{reflected})
and~(\ref{Epm}) into Eq.~(\ref{total-signal}) and then into
Eq.~(\ref{intensity}) and retaining only terms of the first order in
the nanostructure response $E_{\mathrm{S}\pm}$ one finds the TRKR signal
\begin{eqnarray}\label{TRKR}
\Delta I(t)
& = &  \frac{1}{\mu_{0}c}
\im\left[ E_{\mathrm{R}}^{*}(t)E_{\mathrm{S}+}(t)
+E_{\mathrm{R}}(t)E_{\mathrm{S}-}^{*}(t) \right]
\nonumber \\
&= & \frac{1}{2}\nu\omega d \re \left[ 
E_{\mathrm{R}}^{*}(t) \btup\rho(t)\kup e^{-i\varphi}
\right.
\nonumber\\
&&\left. -E_{\mathrm{R}}(t) \btdn\rho(t)\kdn^{*} e^{i\varphi}
 \right]  
\end{eqnarray}
In the above discussion, we have assumed that all the
hole-trion systems evolve under the same conditions. The effect of
inhomogeneity will be treated in Sec.~\ref{sec:inhomo}.
Eq.~(\ref{TRKR}) describes the measured signal in terms of the quantum
state of a nanostructure. This equation
can be used to find the system response without any further
simplifying assumptions based on a numerical simulation of the
open system evolution.

The next step
is to derive the relation between the elements of the density matrix
and the spin polarization 
before the arrival of the probe pulse. This relation can be expressed in an
analytical form \cite{yugova09}
under the assumption that the probe pulse is much shorter
than any relevant time scale of the system dynamics (consistent with
the idea that it is supposed to
probe the instantaneous state of the system). One has to assume also that
the dephasing times of interband coherences are longer than the
pulse duration.

In order to relate the Kerr response to the density matrix formalism
we note that the system state $\rho(t)$, which gives rise to the measured
polarization, is prepared by the probe pulse from the state just before
this pulse, $\rho(\tau^{-})$, where $\tau^{-}$
denotes the time instant just before the arrival of the probe pulse. 
Under conditions stated above, we can completely neglect the system
evolution during the pulse. Then, the system density matrix is
transformed according to $\rho(t)=W(t)\rho(\tau^{-}) W^{\dag}(t)$,
with the unitary operator
\begin{eqnarray*}
W(t) & = &\cos\frac{\Phi_2(t)}{2}\mathbb{I}\\
&&-i\sin\frac{\Phi_2(t)}{2} \left[\left(
\kup\btup+\kdn\btdn\right)e^{i\psi_{2}} +\mathrm{H.c.} \right],
\end{eqnarray*}
where 
\begin{displaymath}
\Phi_{2}(t)=\frac{A_{2}}{\sqrt{2}\hbar}
\int_{-\infty}^{t}d s \eta\left( \frac{s-\tau}{\tau_{p}} \right).
\end{displaymath}

With this time evolution operator one finds for the interband matrix
elements (for $\sigma=\uparrow,\downarrow$)
\begin{eqnarray}
\lefteqn{\left\langle T\sigma|\rho(t)|\sigma\right\rangle=}\nonumber\\ 
&& \cos^{2}\frac{\Phi_{2}(t)}{2} \left\langle T\sigma|
\rho(\tau^{-})|\sigma\right\rangle \nonumber\\
&& +\frac{i}{2}\sin\Phi_{2}(t)
\left[ \left\langle T\sigma|\rho(\tau^{-})|T\sigma\right\rangle
-\left\langle\sigma|\rho(\tau^{-})|\sigma\right\rangle \right]
e^{-i\psi_{2}} \nonumber\\
&& +\sin^{2}\frac{\Phi_{2}(t)}{2}
\left\langle\sigma|\rho(\tau^{-})|T\sigma\right\rangle
e^{-2i\psi_{2}}.
\label{pol-up}
\end{eqnarray}
Let us first assume that the delay time between the pump and the probe
pulse is much longer than the interband dephasing time. (We will come back
to the case when this is not fulfilled in Sec. \ref{sec:spin-opt}). In
this case, the 
interband matrix elements at time $\tau^{-}$ are negligible and we only
have the contributions proportional to $\sin \Phi_2(t)$, i.e., the terms
involving the occupation differences between trion and hole states.
Substituting this into Eq.~(\ref{TRKR}) and using the expression (\ref{reflected})
for the reflected field we find for the TRKR signal
\begin{eqnarray}
\Delta I & = & 
\frac{1}{4\sqrt{2}}rE_{2}\eta\left( \frac{t-\tau}{\tau_{\mathrm{p}}}\right) 
\nu\omega d\sin\varphi \sin\Phi_{2}(t) \nonumber \\
&&\times \left[ \Sigma_{\mathrm{t}}(\tau^{-})-\Sigma_{\mathrm{h}}(\tau^{-})
\right],
\label{TRKR-micro}
\end{eqnarray}
where 
\begin{subequations}
\begin{eqnarray}
\Sigma_{\mathrm{t}}(t)&=&\btup\rho(t)\ktup-\btdn\rho(t)\ktdn,
\label{Sigma-a}\\
\Sigma_{\mathrm{h}}(t)&=&\bup\rho(t)\kup-\bdn\rho(t)\kdn,
\label{Sigma-b}
\end{eqnarray}
\end{subequations}
are trion and hole spin polarizations, respectively.

We neglect here the delay between the field envelopes due to propagation
through the capping layer, which is of the order of 1~fs, that is, much shorter
than the picosecond pulse duration.
The signal described by Eq.~(\ref{TRKR-micro}) is
proportional to the difference of hole and trion spin polarizations just
before the probe pulse. In this way, the TRKR measurement gives access
to the evolution of the spin polarizations in the system.

For a pulsed excitation, $\Delta I$ depends on time. We define the
time-integrated (TI) TRKR signal as
\begin{equation}\label{TI-TRKR}
\Delta I_{\mathrm{TI}}=\int_{-\infty}^{\infty}dt \Delta I(t).
\end{equation}
This quantity is a function of the time delay $\tau$ between the pump
and probe pulses. Since the homodyne response is proportional to the
envelope of the probe pulse, the integration in the above equation is
done over the duration of the probe pulse. 

We note that
\begin{displaymath}
\int_{-\infty}^{\infty} dt \eta\left( \frac{t-\tau}{\tau_{\mathrm{p}}}
\right) \sin\Phi_{2}(t)
=\frac{\sqrt{2}\hbar}{A_{2}}\left[ 1-\cos\alpha_{2} \right],
\end{displaymath}
where 
\begin{displaymath}
\alpha_{2}=\frac{A_{2}}{\sqrt{2}\hbar}
\int_{-\infty}^{\infty}dt \eta\left( \frac{t-\tau}{\tau_{p}} \right)
\end{displaymath}
is the area of the probe pulse. Hence, the integrated detection signal
is (see also Ref.~\onlinecite{yugova09})
\begin{equation}\label{TRKR-TI-micro}
\Delta I_{\mathrm{TI}}=
\nu\hbar\omega\frac{r\sin\varphi}{4(1+r)}
\left( 1-\cos\alpha_{2} \right)
\left[ \Sigma_{\mathrm{t}}(\tau^{-})-\Sigma_{\mathrm{h}}(\tau^{-}) \right]. 
\end{equation}

In the weak pulse limit, the
spin polarizations are proportional to the intensity of the pump
pulse, hence the signal is also proportional to this
intensity. Moreover, it follows directly from
Eq.~(\ref{TRKR-TI-micro}) that in this limit the response is also
proportional to the intensity of the probe pulse.

The quantity $\nu\hbar\omega$ sets the natural energy scale for the
emitted radiation and is equal to the energy the system would emit per
unit area if each hole-trion system generated one photon. Thus,
$\Delta I_{\mathrm{TI}}/(\nu\hbar\omega)$, which is the quantity to be
plotted based on the results of simulations in
Secs.~\ref{sec:spin-opt} and~\ref{sec:results}, corresponds to the
average number of photons per one hole-trion emitter and one
repetition of the experiment.

\section{Hole and trion spin decoherence}
\label{sec:decoh}

In this section we present a detailed analysis of the spin dynamics
based on an analytical solution of the equation of motion
for the density matrix in an idealized situation of coherent optical
driving and fast dephasing of optical coherences. The validity of
these assumptions and the relation between the quantities calculated
here and the actual signal are discussed in Sec.~\ref{sec:spin-opt}. 
We begin with a discussion of dephasing of a single system. Then we
study the effect of inhomogeneity of $g$-factors across the ensemble.

\subsection{Homogeneous dephasing}
\label{sec:homo}

The set of dynamical variables describing the evolution of the system
consists of the trion and hole populations
\begin{equation*}
N_{\mathrm{t}}(t)=\btup\rho(t)\ktup+\btdn\rho(t)\ktdn,\quad
N_{\mathrm{h}}(t)=1-N_{\mathrm{t}}(t),
\end{equation*} 
the trion and hole spin polarizations defined in Eqs.~(\ref{Sigma-a})
and (\ref{Sigma-b}),
as well as trion and hole spin coherences,
\begin{eqnarray*}
X_{\mathrm{t}}(t)&=&\btup\rho(t)\ktdn+\btdn\rho(t)\ktup,\\
Y_{\mathrm{t}}(t)&=&i(\btup\rho(t)\ktdn-\btdn\rho(t)\ktup), \\
X_{\mathrm{h}}(t)&=&\bup\rho(t)\kdn+\bdn\rho(t)\kup,\\
Y_{\mathrm{h}}(t)&=&i(\bup\rho(t)\kdn-\bdn\rho(t)\kup).
\end{eqnarray*}

Initially, all trion variables are zero. The spin of the trapped hole is
in the thermal equilibrium 
state, which, in the basis of the hole spin
eigenstates, is characterized by a spin polarization
\begin{equation*}
p=\bp\rho_{\mathrm{eq}}\kp-\bmm\rho_{\mathrm{eq}}\km=
\tanh\left( \frac{\hbar\wh }{2k_{\mathrm{B}}T} \right),
\end{equation*}
where $\rho_{\mathrm{eq}}$ is the density matrix for the system state at
equilibrium. 
This corresponds to the following initial values for the
dynamical variables of the holes:
\begin{eqnarray*}
N_{\mathrm{h}}|_{t<0}  =  1 , \quad \Sigma_{\mathrm{h}}|_{t<0} = p \cos \phi , \\
X_{\mathrm{h}}|_{t<0} = p \sin \phi , \quad Y_{\mathrm{h}}|_{t<0} = 0.
\end{eqnarray*}

In the present discussion, we assume that the pulse durations are much
shorter than any characteristic 
time scale of the spin dynamics and their action may be approximately treated as
instantaneous. This means that the
excitation is coherent (the case of dephasing times
comparable with pulse durations is discussed in
Sec.~\ref{sec:spin-opt}). We assume also that it is resonant 
(effects of detuning 
for the case of $n$-doped structures and exact Voigt geometry
have been studied in Ref.~\onlinecite{yugova09}).
Then, the effect of the pump 
pulse is to perform the 
rotation $\rho\to V\rho V^{\dag}$, where
\begin{eqnarray*}
V& = & \pdn+\ptdn+\cos\frac{\alpha_{1}}{2}(\pup+\ptup)\\
&&-i\sin\frac{\alpha_{1}}{2}
(\kup\!\btup e^{i\psi_{1}}+\ktup\!\bup e^{-i\psi_{1}}),
\end{eqnarray*}
and
\begin{equation*}
\alpha_{1}=\frac{\tau_{\mathrm{p}}A_{1}}{\hbar}
\int_{-\infty}^{\infty}ds\eta(s)
\end{equation*}
is the pulse area. 
The pulse generates the trion population and depletes 
the hole population accordingly,
\begin{equation*}
N_{\mathrm{t}}(0)=
\sin^{2}\frac{\alpha_{1}}{2}\frac{p\cos\phi+1}{2},\quad
N_{\mathrm{h}}(0)
=1-N_{\mathrm{t}}(0),
\end{equation*}
generates the hole and trion polarization,
\begin{eqnarray*}
\Sigma_{\mathrm{h}}(0) & = & 
\frac{\left( 1+\cos^{2}\frac{\alpha_{1}}{2} \right)p\cos\phi 
-\sin^{2}\frac{\alpha_{1}}{2}}{2},\\
\Sigma_{\mathrm{t}}(0) & = & \sin^{2}\frac{\alpha_{1}}{2}\frac{p\cos\phi+1}{2},
\end{eqnarray*}
and reduces the hole spin coherence which exists at thermal
equilibrium in a tilted magnetic field
\begin{equation*}
X_{\mathrm{h}}(0)=\cos\frac{\alpha_{1}}{2}p\sin\phi.
\end{equation*}
The other dynamical variables remain zero.

The subsequent dynamics of the system is generated by the Zeeman
Hamiltonian $H_{0}$ and by the dissipators $L_{\mathrm{h}}$ and
$L_{\mathrm{rad}}$,
\begin{equation}\label{LvNL}
\dot{\rho}=-\frac{i}{\hbar}[H_{0},\rho]+
L_{\mathrm{h}}[\rho]+L_{\mathrm{r}}[\rho].
\end{equation}
For the occupation, this yields a single decay equation with the
obvious solution
\begin{equation*}
N_{\mathrm{t}}=N_{\mathrm{t}}(0)e^{-\gamma_{1}t},\quad
N_{\mathrm{h}}=1-N_{\mathrm{t}}.
\end{equation*}
When the trion spin dephasing is neglected
the trion variables evolve according to a closed set of three equations,
\begin{eqnarray*}
\dot{\Sigma}_{\mathrm{t}} & = & -\gamma_{1}\Sigma_{\mathrm{t}}
-\wt \sin\theta Y_{\mathrm{t}} \\
\dot{X}_{\mathrm{t}}& = & -\gamma_{1}X_{\mathrm{t}}
+\wt \cos\theta Y_{\mathrm{t}} \\
\dot{Y}_{\mathrm{t}}& = & -\gamma_{1}Y_{\mathrm{t}}
-\wt \cos\theta X_{\mathrm{t}} 
+\wt \sin\theta \Sigma_{\mathrm{t}},
\end{eqnarray*}
where $\wt=g_{\mathrm{t}}\mu_{\mathrm{B}}B/\hbar$ is the trion Larmor frequency.
The solution for the trion spin polarization is easily found to be 
\begin{equation}\label{Sigma-t}
\Sigma_{\mathrm{t}}(t)=\Sigma_{\mathrm{t}}(0)e^{-\gamma_{1}t}
\left[ \cos^{2}\theta+\cos\wt t\sin^{2}\theta \right].
\end{equation}

\begin{figure}[tb]
\begin{center}
\includegraphics[width=42mm]{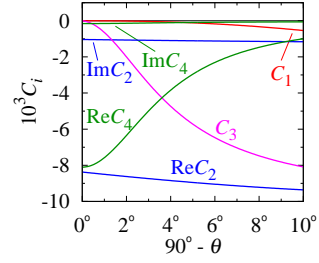}
\end{center}
\caption{\label{fig:C}(Color online) The amplitudes
  of the contributions to 
  the TRKR signal [Eqs.~(\ref{Sigma-t}) and (\ref{Sigma-h})] as a
  function of the tilt angle of the magnetic field. The amplitudes
  $C_{1}$ and $C_{2}$ include both the hole contributions
  [Eq.~(\ref{Sigma-h})] and the corresponding trion contributions from
  Eq.~(\ref{Sigma-t}) with the same time dependence.}
\end{figure}

For the hole variables, the equation of motion (\ref{LvNL}) leads to
the system of equations with
$N_{\mathrm{h}}$ and $\Sigma_{\mathrm{t}}$ acting as source terms,
\begin{eqnarray*}
\dot{\Sigma}_{\mathrm{h}} & = & 
-\left(\kappa_{0}\sin^{2}\phi+\kappa_{1}\frac{1+\cos^{2}\phi}{2}\right)
\tilde{\Sigma}_{\mathrm{h}}\\
&&+\frac{1}{2}\left( \kappa_{0}-\frac{\kappa_{1}}{2} \right) \sin 2\phi 
\tilde{X}_{\mathrm{h}}-\wh \sin\phi Y_{\mathrm{h}}\\
&&-\kappa' \cos\phi N_{\mathrm{h}} +\gamma_{1}\Sigma_{\mathrm{t}},\\
\dot{X}_{\mathrm{h}}& = & 
-\left( \kappa_{0}\cos^{2}\phi- \frac{\kappa_{1}}{2}\cos 2\phi
\right)\tilde{X}_{\mathrm{h}} +\wh \cos\phi Y_{\mathrm{h}}\\
&&+\frac{1}{2}\left(\kappa_{0}-\frac{\kappa_{1}}{2}\right)
\sin 2\phi\tilde{\Sigma}_{\mathrm{h}}
 -\frac{1}{2}\kappa'\sin 2\phi N_{\mathrm{h}}\\
\dot{Y}_{\mathrm{h}}& = & 
-\left[ \kappa_{0}+\frac{\kappa_{1}}{2} \right] Y_{\mathrm{h}}
-\wh \cos\phi \tilde{X}_{\mathrm{h}} +\wh \sin\phi \tilde{\Sigma}_{\mathrm{h}},
\end{eqnarray*}
where $\kappa_{1}=\kappa_{+}+\kappa_{-}$, 
$\kappa'=\kappa_{+}-\kappa_{-}$, and 
\begin{displaymath}
\widetilde{\Sigma}_{\mathrm{h}}=\Sigma_{\mathrm{h}}-p\cos\phi,\quad
\widetilde{X}_{\mathrm{h}}=X_{\mathrm{h}}-p\sin\phi
\end{displaymath}
(the equilibrium values are subtracted).
By the Laplace transform technique, one finds the solution for the
hole spin polarization in the form 
\begin{equation}\label{Sigma-h}
\Sigma_{h}(t)=\sum_{k=1}^{5}C_{k}e^{\lambda_{k}t}+\mathrm{c.c.}
\end{equation}
The exponents $\lambda_{k}$ can be found exactly in a simple form,
while the expressions for the amplitudes $C_{k}$ are rather lengthy,
therefore we use the 
fact that the hole dephasing rates $\kappa_{\pm,0}$ are much smaller
than all the other rates and frequencies and give the formulas for
$C_{k}$ in the leading order, that is, neglecting terms
$O(\kappa_{\pm,0})$. The result is
\begin{eqnarray*}
\lambda_{1}& = & -\gamma_{1},\\
C_{1}&=& -\frac{1}{2}
\frac{\wh^{2}\cos^{2}\phi+\gamma_{1}^{2}}
{\wh^{2}+\gamma_{1}^{2}}\cos^{2}\theta
\Sigma_{\mathrm{t}}(0);\\
\lambda_{2} & = & -i\wt -\gamma_{1}, \\
C_{2} & = &
-\frac{1}{2}\frac{\gamma_{1}}{\gamma_{1}+i\wt }
\frac{\wh^{2}\cos^{2}\phi+(\gamma_{1}+i\wt )^{2}}{
\wh^{2}+(\gamma_{1}+i\wt )^{2}}
\sin^{2}\theta \Sigma_{\mathrm{t}}(0);\\
\lambda_{3} & = & -\kappa_{1},\\
C_{3} & = & -\frac{1}{2}\frac{\wt^{2}}{\wt^{2}+\gamma_{1}^{2}}
\cos^{2}\phi\sin^{2}\theta\Sigma_{\mathrm{t}}(0)
+\frac{1}{4}\sin 2\phi \widetilde{X}_{\mathrm{h}}(0);\\
\lambda_{4} & = & -i\wh-\kappa_{0}-\frac{\kappa_{1}}{2},\\
C_{4} & = & \frac{1}{2}\frac{\gamma_{1}}{\gamma_{1}-i\wh}
\frac{\wt^{2}\cos^{2}\theta+(\gamma_{1}-i\wh)^{2}}{
\wt^{2}+(\gamma_{1}-i\wh)^{2}}\sin^{2}\phi\Sigma_{\mathrm{t}}(0) \\
&&+\frac{1}{2}\sin^{2}\phi\widetilde{\Sigma}_{\mathrm{h}}(0) 
-\frac{1}{4}\sin 2\phi \widetilde{X}_{\mathrm{h}}(0);\\
\lambda_{5}& = & 0, \\
C_{5} & = & p\cos\phi.
\end{eqnarray*}
The values of the amplitudes $C_{i}$ as a function of the orientation
of the magnetic field for the parameters assumed in
this paper (Tab.~\ref{tab:params}) are plotted in Fig.~\ref{fig:C}.

\begin{table}[tb]
  \centering
\begin{tabular}{ll}
\hline
Electron g factor & $g_{e}=0.26$ \\
Hole g factor & \\
\ \ \  -- axial & $g_{||}=0.6$ \\
\ \ \  -- in plane & $g_{\bot}=0.04$ \\
Trion recombination time & $1/\gamma_{1}=50$~ps \\
Refractive index & $n=3.44$ \\
Pulse duration (pump \& probe) & $\tau_{\mathrm{p}}=1$~ps \\
Pulse amplitude & \\
\ \ \  -- pump & $dE_{1}=0.5$~meV \\
\ \ \  -- probe & $dE_{2}=0.1$~meV \\
Temperature & $T=1.6$~K \\
Magnetic field & $B=7$~T \\
\hline
\end{tabular}
\caption{\label{tab:params}System parameters which are fixed
  throughout the paper. The hole $g$ factors are taken as for a 6~nm
  thick quantum well \cite{marie99,snelling92}. Parameters correspond
  to an AlGaAs structure \cite{syperek07}.}
\end{table}

According to Eqs.~(\ref{Sigma-t}) and (\ref{Sigma-h}), there are three
kinds of contributions to the total spin polarization 
$\Sigma_{\mathrm{h}}-\Sigma_{\mathrm{t}}$. The constant one,
$(C_{5},\lambda_{5})$, corresponds to the equilibrium 
spin polarization. Exponentially decaying contributions, given by the first term
in Eq.~(\ref{Sigma-t}) and by the 1st and 3rd term in
Eq.~(\ref{Sigma-h}), originate from the decay of the spin population with
respect to the respective quantization axes. Since we assumed that the
trion spin lifetime is limited by the recombination time, the trion
spin population decays with the recombination rate $\gamma_{1}$. The
hole population decays with the spin relaxation rate
$\kappa_{1}$. Since only spin polarization along the growth axis is
relevant in the optical measurement, these contributions vanish when
the spin quantization axis is perpendicular to the structure symmetry
axis, that is, $\cos\phi=0$ and $\cos\theta=0$ for the hole and trion
contributions, respectively. It should be noted that due to the
strong anisotropy of the hole $g$ factor, the out-of-plane component
of the hole spin is large already in a slightly tilted magnetic
field. Therefore, the occupations of the
Zeeman levels and their thermalization affect the optical response
already in slightly tilted fields.
The second term in
Eq.~(\ref{Sigma-t}) and the 2nd and 4th terms in Eq.~(\ref{Sigma-h}) reflect
the spin precession around the quantization axis. This precession
affects the optically detected spin polarization only if the
quantization axis is tilted with respect to the structure axis, that
is, $\sin\phi\neq 0$ and $\sin\theta\neq 0$.

\subsection{Inhomogeneous effects}
\label{sec:inhomo}

If the measured signal originates form an ensemble of emitters, it
becomes dephased due to variation of system parameters across the
ensemble. In our case, non-uniformity of $g$-factors makes the
individual spins precess with various rates, which destroys the
overall spin polarization. 

We assume that the number of systems in the ensemble is sufficient to
describe the distribution of $g$-factors by a continuous distribution
function. We neglect possible variation of the quantization axis. It
is convenient to describe the inhomogeneity in terms of the trion and
hole precession frequencies $\wt$ and $\wh$, for which we assume
Gaussian distributions
\begin{equation}\label{ditrib}
f_{i}(\tilde{\omega}_{i})=\frac{1}{\sqrt{2\pi}\sigma_{i}}
e^{-\frac{(\tilde{\omega}_{i}-\omega_{i})^{2}}{2\sigma_{i}^{2}}},\quad
i=\mathrm{t,h},
\end{equation}
where $\wt$, $\wh$ now become the central frequencies of the
corresponding distributions.
We assume that $\sigma_{i}\ll\omega_{i}$, so that a variation of the
amplitudes $C_{k}$ in Eq.~(\ref{Sigma-h}) can be neglected.
Then, upon averaging with the distribution functions (\ref{ditrib}), the trion and hole
spin polarizations [Eqs.~(\ref{Sigma-t}) and (\ref{Sigma-h})] become
\begin{eqnarray}\label{Sigma-t-inh}
\Sigma_{\mathrm{t}}(t) & = & \Sigma_{\mathrm{t}}(0)e^{-\gamma_{1}t}
\left[ \cos^{2}\theta+e^{-\sigma_{\mathrm{t}}^{2}t^{2}/2}\cos\wt
  t\sin^{2}\theta \right],\\
\label{Sigma-h-inh}
\Sigma_{h}(t) & = & \sum_{k=1}^{5}C_{k}e^{f_{k}(t)}+c.c.,
\end{eqnarray}
where the amplitudes $C_{k}$ are the same as in Eq.~(\ref{Sigma-h}), 
$f_{k}(t)=\lambda_{k}t$ for $k=1,3,5$, 
$f_{2}(t)=-i\omega_{\mathrm{t}}t-\gamma_{1}t-\sigma_{\mathrm{t}}^{2}t^{2}/2$, and
$f_{4}(t)=-i\omega_{\mathrm{h}}t-(\kappa_{0}+\kappa_{1}/2)t-\sigma_{\mathrm{h}}^{2}t^{2}/2$.
As usual, the exponential decay of a single system is replaced by a
Gaussian one if the dispersion of frequencies is larger than the
homogeneous dephasing rates.

\section{Spin polarization and TRKR response}
\label{sec:spin-opt}

The discussion presented in the previous sections was based on some
simplifying assumptions. On the one hand, in our discussion of the TRKR
response in Sec.~\ref{sec:signal}, we concentrated on delay times longer than the
interband dephasing time. Therefore we did not discuss the contributions to
the interband polarization resulting from the interband coherences
created by the pump pulse. In quantum wells, such interband coherences
vanish very quickly but in self-assembled quantum dots their lifetime may be
limited only by the recombination time, which is of the order of a
nanosecond \cite{langbein04,bayer02}. On the other hand, the
derivation of the analytical 
formulas in Sec.~\ref{sec:decoh}, as well as of the relation between the spin
polarization and the TRKR signal, is based on the assumption of coherent
excitation. This, in turn, requires the coherence time to be long enough
and the coherence assumption breaks if dephasing of the optical coherences
is fast. In this section, we will deal with these issues.

In order to model the full optical response of the system we will
numerically solve the evolution equation
\begin{equation}\label{evol-num}
\dot{\rho}=-\frac{i}{\hbar}\left[ H_{0}+H_{\mathrm{las}},\rho \right]
+L_{\mathrm{h}}[\rho]+L_{\mathrm{rad}}[\rho],
\end{equation}
calculate the optical signal according to Eq.~(\ref{TRKR}), and
integrate the result according to Eq.~(\ref{TI-TRKR}).
Some of the parameters will be kept constant for all the results
presented in this paper. The
values of these fixed parameters (roughly correspopnding to an AlGaAs QW
system similar to that studied in Ref.~\onlinecite{syperek07}) are
collected in Tab.~\ref{tab:params}. 
The trion Larmor frequency is
$\omega_{\mathrm{t}}=0.16$~ps$^{-1}$. The pulse amplitudes chosen here
correspond to 
the pulse areas $\alpha_{1}=0.12\pi$ and $\alpha_{2}=0.016\pi$ for the
pump and probe pulse, respectively. These values assure that the
optical excitation is well in the linear regime, so that varying the pulse
areas leads only to uniform rescaling of the signal intensity
proportionally to the pulse intensities, that is, to $\alpha_{1}^{2}$
and $\alpha_{2}^{2}$.

First, we will discuss the additional contributions in the case of delay
times shorter or of the same order as the interband dephasing time. In this
case, also the interband terms at time $\tau^{-}$ in Eq.~(\ref{pol-up})
contribute to the total interband matrix elements at
time $t$ and thus to the emitted radiation. However, we notice that the
phase of the second pulse $\psi_2$ enters differently in the three terms.
Keeping in mind that the interband coherences created by the first pulse
carry a phase factor $e^{-i\psi_1}$ (in the case of 
$\btup\rho(\tau^{-})\kup$) and $e^{i\psi_1}$ (in the case of 
$\bup\rho(\tau^{-})\ktup$), the total phases are $\psi_1$ for the
first term, $\psi_2$ for the second term, and $2\psi_2-\psi_1$ for the
third term. Hence, only the second term holds a fixed phase relation
with the reflected beam and can produce a non-vanishing homodyne
signal if the relative phase between the pulses is random. Moreover,
the two exciting laser pulses are usually applied to the sample 
at slightly different directions $\bm{k}_1$ and $\bm{k}_2$. Then, for
an extended system (ensemble of emitters),
also the emitted radiation originating from the three contributions has
different directions. The first one is emitted in the direction
$\bar{\bm{k}}_1$ of the reflected pump pulse, the second one in the
direction of the reflected probe pulse $\bar{\bm{k}}_2$ and the third
one in the background-free reflected four-wave mixing direction
$2\bar{\bm{k}}_2 -\bar{\bm{k}}_1$. Thus we find that the interband
coherences resulting from the pump pulse excitation indeed give rise to
emitted signals, but they do not contribute to the TRKR signal
[Eq.~(\ref{TRKR})]. 
The only exception would be the case of temporally overlapping pump and
probe pulses, where the actions of pump and probe pulses cannot be treated
separately, and the case of perfectly aligned, phase-locked pulses where
the interference between the interband coherence from the first pulse and
the second pulse gives rise to strong coherent control oscillations. Both
cases, however, are not relevant for the purpose of this paper.

\begin{figure}[tb]
\begin{center}
\includegraphics[width=85mm]{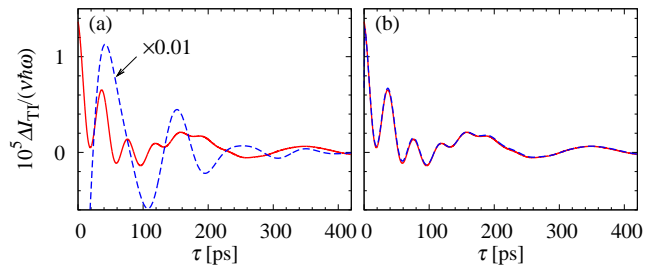}
\end{center}
\caption{\label{fig:compar}(Color online) Comparison between the analytical result
  for the spin polarization (red solid lines) and the simulated signal
  (dashed blue lines) for
  $\theta=86^{\circ}$, $1/\kappa_{1}=100$~ps, $\gamma_{0}=0$ (slow
  optical dephasing). In (a),
  the hypothetical detection signal corresponding to a fixed phase relation
  between the pump and probe pulses is shown; in (b), the simulated
  signal has been averaged over the phase of the probe pulse.}
\end{figure}

The numerical solution for the system evolution according to
Eq.~(\ref{evol-num}) does not involve any simplifying 
assumptions, apart from the resonant excitation condition. However, in
this model, the geometrical relations between various directions in
which the signals are emitted are not taken into account. Moreover,
in a single simulation run, the relative phase between the pulses is
fixed. Therefore, the calculated optical response contains both the
TRKR signal and the coherent components. A comparison of the simulated
signal to the spin polarization in this case, shown in
Fig.~\ref{fig:compar}(a), reveals that the former not only differs by
orders of magnitude but also is uncorrelated to the latter. This is
the case even for delay times a few times longer than the trion
relaxation time since the coherent contribution belongs to a lower
order of the optical response and is many orders of magnitude stronger
in the weak excitation limit. The
coherent artefacts can be eliminated from the simulation result by
simply averaging the results obtained with opposite signs of the 
probe amplitude. If this is done, the simulated signal agrees with the
analytical formulas, as expected [see Fig.~\ref{fig:compar}(b)]. This
confirms that the approximations made in the analytical solution do
not noticeably affect the result.

\begin{figure}[tb]
\begin{center}
\includegraphics[width=85mm]{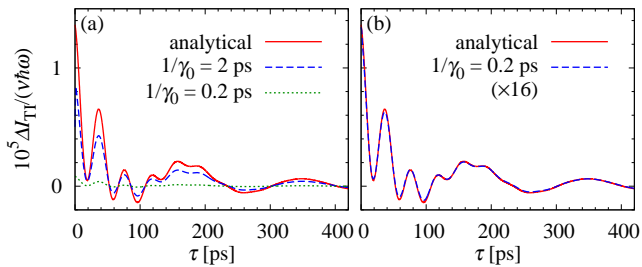}
\end{center}
\caption{\label{fig:compar2}(Color online) (a) Comparison between the analytical result
  for the spin polarization (red solid lines) and the detection signal
  (dashed blue lines) for
  $\theta=86^{\circ}$, $1/\kappa_{1}=100$~ps, $\gamma_{0}\neq 0$, as
  shown (fast optical dephasing case). (b) The simulated signal has
  been rescaled up by a factor of 16 to show that its shape exactly
  follows the evolution predicted by the analytical formulas.}
\end{figure}

In the opposite case of strong interband dephasing, the analytical formulas are
no longer valid. In Fig.~\ref{fig:compar2}(a) we compare the analytical
result (red solid line) with the simulated signal for two values of
the dephasing rate $\gamma_{0}$ which describes additional pure
dephasing of the optical coherence (beyond that associated with the
radiative decay the rate of which, $\gamma_{1}$ is fixed throughout the
paper). As the dephasing time becomes comparable with the pulse
duration, the signal is quenched due to the reduced efficiency of optical
pumping and probing. We note, however, that this quenching is
uniform, that is, it does not modify the shape of the pulse. This is
clear from Fig.~\ref{fig:compar2}(b), where the simulated response for
$\gamma_{0}=5$~ps$^{-1}$ has been multiplied by a factor of 16. Upon
this rescaling, the simulated signal matches the analytically
calculated one almost exactly.

Thus, we have established the relation between the evolution of spin
polarization in the system and the form of the TRKR response. It turns
out that both the simulated (or measured) signal and the analytical
formula can yield consistent, correct information on the spin
evolution. One has to eliminate the coherent
polarization contributions from the calculated optical response in
the slow dephasing case and the analytical formulas uniformly
overestimate the signal in the case of fast optical dephasing. 

\section{Results}
\label{sec:results}

In this Section, we discuss the evolution of the spin polarization,
based on the analytical solution to the equations of motion derived in
Sec.~\ref{sec:decoh}. In all the simulations presented below, we set
$\gamma_{0}=0$ (hence, the term ``pure dephasing'' will always refer
to the pure dephasing of spin states, described by the parameter
$\kappa_{0}$).  

\begin{figure}[tb]
\begin{center}
\unitlength 1mm
\begin{picture}(85,30)(0,5)
\put(0,0){\resizebox{85mm}{!}{\includegraphics{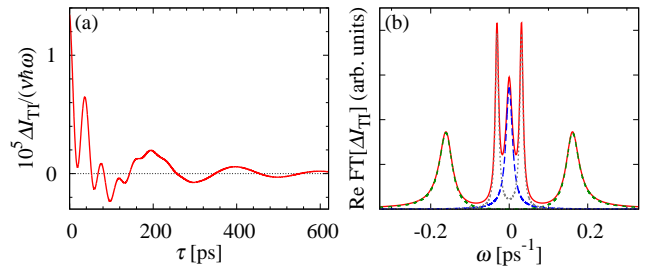}}}
\end{picture}
\end{center}
\caption{\label{fig:general}(Color online)  (a) The TRKR signal for
  $\theta=87^{\circ}$, $1/\kappa_{1}=100$~ps,
  $\kappa_{0}=\sigma_{\mathrm{h}}=\sigma_{\mathrm{t}}=0$. (b) The
  real part of the Fourier transform of the signal (red solid line)
  and the contributions from the trion precession (green short-dashed
  line), hole precession (grey dotted line) and hole spin relaxation
  (blue long-dashed line). The contribution from the exciton spin
  relaxation is invisible on this scale.} 
\end{figure}

In Fig.~\ref{fig:general}(a), we show the evolution of the spin
polarization for a certain 
set of parameters. The signal appears to be dominated by two
oscillating components. As discussed in Sec.~\ref{sec:homo}, the
short-period one corresponds to the trion 
precession with the frequency $\omega_{\mathrm{t}}$. This contribution
is damped with the rate $\gamma_{1}$ due to the finite trion life
time. The other oscillating contribution originates from the hole
precession and is damped with the total hole spin dephasing rate
$\kappa_{1}/2+\kappa_{0}$, reflecting the decay of the
hole spin coherence (transverse dephasing). 
Two other contributions, which are less evident in the plot, have a
non-oscillating character and reflect the spin relaxation leading to thermalization
between the Zeeman 
eigenstates (longitudinal decoherence). 
The presence of these parts of the signal becomes easily visible in
the form of a central line in the Fourier transform of the TRKR
response, shown in Fig.~\ref{fig:general}(b). We plot here also the individual
contributions following from Eqs.~(\ref{Sigma-t}) and (\ref{Sigma-h}).

The two most interesting aspects of the TRKR response
are the dependence of the signal on the tilt angle between the
magnetic field and structure plane and the effect of various dephasing
types (spin relaxation, pure dephasing, inhomogeneous dephasing). In
the following subsections, we
start our analysis with the angle dependence and later proceed to the
role of various dephasing contributions.

\subsection{Tilt angle dependence}

Due to the strong anisotropy of the hole $g$ factor, the TRKR signal shows
a very strong dependence on the angle at which the magnetic field is
tilted off the system plane. The
quantization axis of the hole spin is far from the plane even for
small tilting angles. Therefore, the form
of the signal 
changes strongly when $\theta$ is varied in the range of a few degrees
from the in-plane orientation.

\begin{figure}[tb]
\begin{center}
\includegraphics[width=85mm]{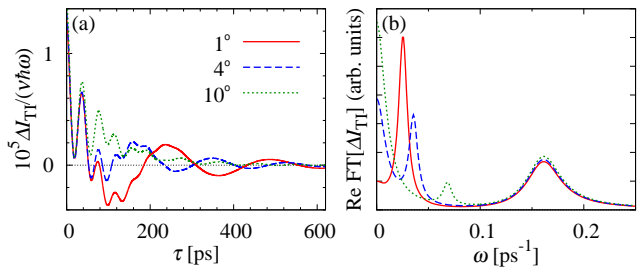}
\end{center}
\caption{\label{fig:tilt}(Color online) (a) The TRKR signal for the three tilt angles
  $90^{\circ}-\theta$ 
  as shown, $1/\kappa_{1}=100$~ps and
  $\kappa_{0}=\sigma_{\mathrm{h}}=\sigma_{\mathrm{t}}=0$. (b) The
  real part of the Fourier transform of the three signals (line coding
  as in (a)). Only the positive frequency part of the symmetric
  spectrum is shown.}
\end{figure}

In Fig.~\ref{fig:tilt}(a) we show the TI-TRKR response 
for three different tilt angles $90^{\circ}-\theta$ between the magnetic field
and the structure plane. In these calculations, we keep the same value
of $\kappa_{1}$ for all angles even though the Zeeman splitting
changes, which is to some extent artificial. A correct dependence
would follow from the detailed modeling of a specific decoherence
channel, which is beyond the scope of the present general description.
In all the cases shown in Fig.~\ref{fig:tilt}(a), one can clearly see
a similar contribution from the trion 
Larmor precession. However, the hole contributions are
different. The
amplitude of the oscillations decreases as the field is tilted more
off the plane. At $90^{\circ}-\theta=10^{\circ}$, the hole
contribution is dominated by a monotonous decay superposed on the
trion oscillations. The reason for this is clear: For the $1^{\circ}$
tilt, the hole precesses around an almost in-plane quantization axis
(oriented at about $15^{\circ}$ off-plane). Such a precession leads to
a strong variation of the perpendicular component of the spin, while
the thermalization of the spin occupations is associated mostly with
the optically irrelevant decay of the in-plane component. On the
contrary, according to
Eq.~(\ref{q-axis}), at $90^{\circ}-\theta=10^{\circ}$ the
hole spin quantization axis is close to perpendicular
($90^{\circ}-\phi=70^{\circ}$). The precession then takes place mostly
in the plane, while the spin population decay affects the perpendicular spin
polarization and is visible in the experiment. 

This qualitative difference in the system evolution is
visible even more clearly in Fig.~\ref{fig:tilt}(b), where we plot the
real part of the Fourier transform of the TI-TRKR signals shown in
Fig.~\ref{fig:tilt}(a). Three characteristic features are visible in
this spectrum. Starting from the right, the broad one at
$\omega=0.16$~ps$^{-1}$ corresponds to the trion precession. The
orientation of the magnetic field does not affect the position of this
feature because the trion (electron) $g$ factor is
isotropic. Moreover, for the narrow range of tilt angles considered
here, the effect on the amplitude of the trion oscillations is very
small. Therefore, this feature is almost insensitive to the
orientation of the field in the considered range. The second feature
moves from 
$\omega=0.025$~ps$^{-1}$ at $90^{\circ}-\theta=1^{\circ}$ to 
$\omega=0.07$~ps$^{-1}$ at $90^{\circ}-\theta=10^{\circ}$ and looses
its amplitude. It corresponds to the hole precession. The frequency
shift is obviously due to the growing contribution of the large axial
component of the hole $g$ factor as the field is tilted off the
plane. The decrease in amplitude corresponds to the fast reorientation of
the hole spin quantization axis, which leads to reduced contribution
of the hole precession to the optical signal. The third feature is the
central line, corresponding to the exponential decay of the hole spin
population. As the magnetic field is oriented more off-plane, the
contribution of this process to the spin polarization grows and this
feature becomes stronger.

\subsection{Dephasing}

\begin{figure}[tb]
\begin{center}
\includegraphics[width=85mm]{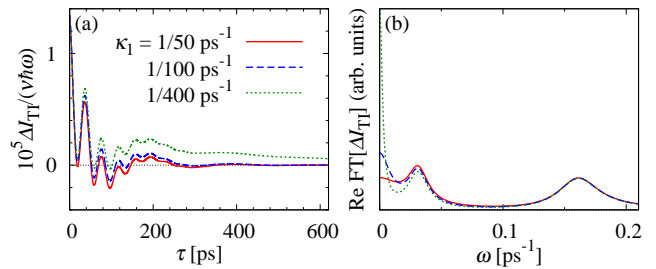}
\end{center}
\caption{\label{fig:homo}(Color online) Comparison of the TRKR
response (a) and its Fourier transform (b) for different contributions
from the pure dephasing as shown 
(the line coding is the same in both panels). The value of
$\kappa_{0}$ is adjusted so that
$\kappa_{1}/2+\kappa_{0}=1/100$~ps$^{-1}$ in all three cases. Here
$90^{\circ}-\theta=3^{\circ}$, $\sigma_{\mathrm{t}}=\sigma_{\mathrm{h}}=0$.} 
\end{figure}

Another interesting feature observed in the simulation is a different
dependence of the decay time of two hole-related components on the two
hole decoherence rates $\kappa_{0}$ and $\kappa_{1}$. This is
visible in Fig.~\ref{fig:homo}, where we fix the precession
damping rate $\kappa_{1}/2+\kappa_{0}$ and change the relative
contributions from the spin relaxation ($\kappa_{1}$) and the
additional pure dephasing ($\kappa_{0}$). In the time-resolved picture
[Fig.~\ref{fig:homo}(a)], the differences are not particularly
characteristic, except for the long exponential tail which develops as
the spin relaxation becomes very slow. Much more pronounced
differences can be noticed in the Fourier transform
[Fig.~\ref{fig:homo}(b)]. As the parameter modification affects only
the hole dynamics, the trion feature remains unchanged. Moreover,
since we fixed the total dephasing time of the hole precession, the
feature at the hole Larmor frequency, $\omega=\wh$, changes very
little (only due to the change in the tails of the neighboring
zero-frequency feature). On the contrary, the central line changes
very strongly. As the lifetime of the spin population becomes longer,
this line gets narrower, with the line area remaining constant. It
seems, therefore, that the spectral components of the TRKR signal in a
tilted magnetic field carry useful information on the relative
strength of different contributions to spin dephasing.

\begin{figure}[tb]
\begin{center}
\includegraphics[width=85mm]{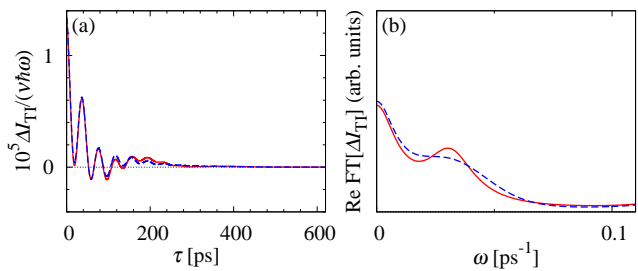}
\end{center}
\caption{\label{fig:inhomo}(Color online) Comparison of the TRKR
response (a) and its Fourier transform (b) for homogeneous and
inhomogeneous dephasing. Only the frequency range relevant to hole
dynamics is plotted in (b). Red solid lines: homogeneous pure dephasing,
$\kappa_{0}=3/400$~ps$^{-1}$, $\sigma_{\mathrm{h}}=0$. Blue dashed
lines: inhomogeneous dephasing, 
$\kappa_{0}=0$, $\sigma_{\mathrm{h}}=0.0147$~ps$^{-1}$. In both cases,  
$90^{\circ}-\theta=3^{\circ}$, $\kappa_{1}=1/100$~ps$^{-1}$.} 
\end{figure}

Another source of damping of the observed spin precession oscillations
in the case of an ensemble measurement is the
inhomogeneity of Larmor frequencies due to a variation of $g$ factors of
the individual emitters in the ensemble. Obviously, only the
precession-related contributions are sensitive to the inhomogeneity
effects. Indeed, as follows from
Eqs.~(\ref{Sigma-t-inh}) and (\ref{Sigma-h-inh}) inhomogeneity
affects the oscillating contributions, while the exponentially
decaying ones remain unaffected. Since the main concern in the present
work is the hole spin evolution, we will restrict the discussion to
the case of $\sigma_{\mathrm{t}}=0$. In Fig.~\ref{fig:inhomo}, we show
the evolution of 
the TRKR signal and the corresponding spectrum for a fixed value of
the hole spin relaxation rate $\kappa_{1}$ with the dephasing
contribution dominated by the homogeneous pure dephasing (red solid
lines) and by $g$ factor inhomogeneity (blue dashed lines). The
parameters $\kappa_{0}$ and $\sigma_{\mathrm{h}}$ are chosen such that
the full width at half maximum of the damping envelope (in the time
domain) is the same in the two cases. Again, there is only a minor
change in the time-resolved signal
[Fig.~\ref{fig:inhomo}(a)]. However, the shape of the spectral feature
corresponding to the hole spin precession changes from Lorentzian to
Gaussian. This change may be characteristic enough to discriminate the
homogeneous vs. inhomogeneous dephasing in a real measurement data.

\section{Discussion and conclusion}
\label{sec:discussion}

We have developed a complete theory of the time-resolved Kerr rotation
experiment for a system of trapped holes in tilted magnetic
fields. The theory is applicable to 
quantum dots or weak trapping centers in quantum wells. In our
approach, we adopted a general description of hole spin relaxation and
dephasing in the Markov limit, based on the Lindblad equation for the
open system dynamics. Spin dephasing is a rather slow process so that
the Markov approximation should work well for this problem
and our approach can be expected to cover a wide range of physical
effects in a way which is independent of the exact microscopic
mechanisms. One should note, however, that there are dephasing
mechanisms that do not admit a Markov approximation of this kind. The
most important example of this class is a spin-environment coupling via
Heisenberg-like (spin-spin) interaction Hamiltonian. 

Our analysis shows that the hole spin dephasing consists actually of
two processes the relative contribution of which depends on the tilt
angle between the magnetic field and the structure plane, with an
important role played by the strong anisotropy of the hole $g$
factor. These two processes are relaxation between the Zeeman states
(occupation thermalization), which dominates the optical response when
the quantization 
axis is close to perpendicular to the plane (aligned with the
structure axis), and dephasing of coherences between the spin states,
which contributes mostly when the quantization axis is close to the
plane. It should be kept in mind that the hole spin quantization axis
is always 
much closer to perpendicular than the magnetic field orientation.

As both these dephasing contributions are marked in the optical signal
for a slightly tilted field (a few degree) a single set of
experimental data conveys, in principle, the full information on the
spin dephasing. Extracting this information is not straightforward at
least for three reasons. First, the Larmor frequencies are not much
higher than dephasing rates and the spectral features related to these
two dynamical contributions overlap rather strongly. Second, the
coefficients of Eq.~(\ref{Sigma-h}) are complex and the features are
not purely Lorentzian. Third, in the case of an ensemble experiment,
inhomogeneous dephasing can dominate the intrinsic one. On the other
hand, Eqs.~(\ref{Sigma-t}) and~(\ref{Sigma-h}) provide analytical
formulas for the spin polarization. As we have shown in
Sec.~\ref{sec:signal}, this spin polarization is identical with the
measured signal (up to uniform rescaling). Then, the formulas provided
by our theory 
can be used to fit the experimental data with just a few parameters,
which might allow one to extract all the relevant decoherence
rates. Moreover, in the present paper, we have used parameter values
which correspond to a quantum well system, where the dephasing is
rather strong. In quantum dots, where spin coherence times are much
longer, the signal should show much more pronounced and separated
features, which can make the analysis much easier. Finally, as shown
in Fig.~\ref{fig:C}, the imaginary parts of the amplitudes
are relatively small, so even a rough line width estimate based on the
Fourier spectrum of the time-resolved signal could yield reasonable
information on the decoherence rates.

\begin{acknowledgments}
The authors are grateful to M. Syperek for discussions.  
This work was supported by Grant No. N N202 1336 33 of the Polish
MNiSW and by a Research Group Linkage Project of the Alexandr von
Humboldt Foundation.
\end{acknowledgments}

\appendix

\section{The Lindblad equation for the spin dephasing}
\label{app:lindblad}

In this Appendix, we derive the general Lindblad equation which
governs the dissipative evolution of the density matrix of a trapped
hole in the Markov limit (an analogous equation can be written for the
trion spin). 

Any observable related to the two-level spin system can be
written as a combination of Pauli matrices $\sigma_{\pm,0}$ acting on
the Hilbert space of hole spin states and written in the basis of
Zeeman eigenstates for a given orientation of the magnetic
field. Therefore, we can write the general Hamiltonian for the
system--reservoir interaction in the form
\begin{equation}
H_{\mathrm{int}}=\sum_{l=\pm,0}\sigma_{l}R_{l},
\label{Hint}
\end{equation}
where $R_{l}$ are certain operators on the Hilbert space of the
reservoir and $R_{+}=R^{\dag}_{-}$. 

One starts with the exact equation of motion for the reduced density
matrix $\tilde{\rho}$ of the hole spin in the interaction picture
\begin{equation}
\frac{d\tilde{\rho}(t)}{dt}
=-\frac{1}{\hbar^{2}}\int_{t_{0}}^{t}d\tau\tr_{\mathrm{R}}
\left[ H_{\mathrm{int}}(t),\left[ H_{\mathrm{int}}(\tau),
\tilde{\varrho}(t) \right]  \right],
\label{evol-exact}
\end{equation}
where $\tilde{\varrho}(t)$ is the density matrix of the total system,
$\tr_{\mathrm{R}}$ denotes partial trace with respect to the reservoir
degrees of freedom, and $t_{0}$ is the initial time of the evolution.

Let us denote the reservoir
memory time by $\tau_{\mathrm{mem}}$.  The Markov approximation is based
on three assumptions\cite{breuer02}: (1) the time $t$ of interest is
much longer than $\tau_{\mathrm{mem}}$; (2) the change of
the system state (in the interaction picture) is small over the time
$\tau_{\mathrm{mem}}$; (3) the relaxation of the reservoir to its
thermal equilibrium is fast compared to the rate with which it is
excited by the system evolution, so that the total density matrix of
the system can be written in a product form, with the reservoir at
equilibrium. Eq.~(\ref{evol-exact}) can be then approximated as
\begin{eqnarray}
\lefteqn{\frac{d\tilde{\rho}(t)}{dt}=}\nonumber\\
&&-\frac{1}{\hbar^{2}}\int_{0}^{\infty}ds\tr_{\mathrm{R}}
\left[ H_{\mathrm{int}}(t),\left[ H_{\mathrm{int}}(t-s),
\tilde{\rho}(t)\otimes \rho_{\mathrm{R}} \right]  \right],
\label{evol-mark}
\end{eqnarray}
where $\rho_{\mathrm{R}}$ is the thermal equilibrium density matrix
of the reservoir. 

In the interaction picture, we denote the reservoir operators $R_{l}$ 
by $R_{l}(t)$ and write the hole spin Pauli matrices as 
$\sigma_{l}(t)=\sigma_{l}e^{-i\omega_{l}t}$, where
$\omega_{-}=-\omega_{+}=\omega_{\mathrm{h}}$ and $\omega_{0}=0$. We
define the reservoir spectral densities
\begin{equation}
R_{lj}(\omega)=\frac{1}{2\pi\hbar^{2}}\int dt e^{i\omega t}
\left\langle R_{l}(t)R_{j} \right\rangle,
\label{spdens}
\end{equation}
where 
$\left\langle R_{l}(t)R_{j} \right\rangle=
\tr_{\mathrm{R}}\rho_{\mathrm{R}}R_{l}(t)R_{j}$.
With this definitions, transforming Eq.~(\ref{Hint}) to the
interaction picture and substituting into Eq.~(\ref{evol-mark}) we get
\begin{eqnarray*}
\frac{d\tilde{\rho}}{dt} & = &
-\sum_{lj}e^{-i(\omega_{l}+\omega_{j})t}\int d\omega R_{lj}(\omega)\\
&&\times\left[ 
\left( \sigma_{l}\sigma_{j}\tilde{\rho}-\sigma_{j}\tilde{\rho}\sigma_{l}
\right) \int_{0}^{\infty}dse^{i(\omega_{j}-\omega)t}s \right.\\
&&\left.+\left( \tilde{\rho}\sigma_{l}\sigma_{j}-\sigma_{j}\tilde{\rho}\sigma_{l}
\right) \int_{0}^{\infty}dse^{i(\omega_{l}+\omega)t}s
 \right]. 
\end{eqnarray*}

In the next step, we use the identity
\begin{displaymath}
\int_{0}^{\infty}ds e^{\pm i\Omega s}
=\pi\delta(\Omega)\pm i\mathcal{P}\frac{1}{\Omega},
\end{displaymath}
where $\mathcal{P}$ denotes the principal value. Moreover, we note
that the terms with $\omega_{l}+\omega_{j}\neq 0$ oscillate quickly in
time and do not contribute considerably to the evolution of the
density matrix. We can thus write
\begin{eqnarray}
\frac{d\tilde{\rho}(t)}{dt}
& = & 2\pi\sum_{lj}\tilde{\delta}_{lj}R_{lj}(\omega_{j})\left( 
\sigma_{j}\tilde{\rho}\sigma_{l}
-\frac{1}{2}\left\{  \sigma_{l}\sigma_{j},\tilde{\rho} \right\}_{+}
 \right) \nonumber \\
&&-\frac{i}{\hbar}[h,\tilde{\rho}],
\label{evol-final}
\end{eqnarray}
where
\begin{displaymath}
h=\hbar\sum_{lj}\tilde{\delta}_{lj}\mathcal{P}\int d\omega
\frac{R_{lj}(\omega)}{\omega_{j}-\omega}\sigma_{l}\sigma_{j},
\end{displaymath}
$\tilde{\delta}_{lj}=1$ if and only if $\omega_{l}+\omega_{j}=0$,
$\{A,B\}_{+}=AB+BA$ and $[,]$ denotes the commutator. 

The second part of the right-hand side of Eq.~(\ref{evol-final}),
containing the commutator, is a correction to the unitary evolution
due to environment-induced level shifts. These effects are very weak
and amount only to a small renormalization of the $g$-factor. We
will, therefore, disregard this term. Of interest to us is the first
term, describing the dissipative impact of the environment. It is
clear that, irrespective of the nature of the reservoir, the dephasing
in the Markov limit, in a given experimental situation, is completely
described by three rates,
\begin{equation}
\kappa_{\pm}=2\pi R_{\pm\mp}(\omega_{\mathrm{h}}),\quad
\kappa_{0}=4\pi R_{00}(0).\label{kappa-R}
\end{equation}
However, using Eq.~(\ref{spdens}), it can be shown that 
$R_{-+}(-\omega)=e^{-\hbar\omega/(k_{\mathrm{B}}T)}R_{+-}(\omega)$,
where $k_{\mathrm{B}}$ is the Boltzmann constant and $T$ is the
temperature. Hence, the number of dephasing parameters reduces to
two. These two dephasing rates are related to
the longitudinal and transverse dephasing times $T_{1}$ and $T_{2}$
(with respect to the quantization axis) by the usual formulas
\begin{displaymath}
T_{1}=\frac{1}{\kappa_{+}+\kappa_{-}},\quad
T_{2}=\frac{1}{\kappa_{0}+1/(2T_{1})}.
\end{displaymath}


\end{document}